\documentclass[showpacs,amsmath,amssymb,twocolumn]{revtex4}
\usepackage{graphicx}
\usepackage{dcolumn}
\usepackage{bm}

\usepackage{color}
\usepackage{ulem}

\begin{document}
\title{Rapidity Dependence of $J/\psi$ Production at RHIC and LHC}
\author{Yunpeng Liu$^1$}
\author{Zhen Qu$^1$}
\author{Nu Xu$^2$}
\author{Pengfei Zhuang$^1$}
\affiliation{$^1$Physics Department, Tsinghua University, Beijing
100084, China\\ $^2$Nuclear Science Division, Lawrence Berkeley
National Laboratory, Berkeley, California 94720, USA}

\begin{abstract}
The motion of charmonium in heavy ion collisions is described by a
three dimensional transport equation with initial production and
continuous regeneration in hot medium. The observation of
apparently stronger $J/\psi$ suppression at forward rapidity
compared to that at midrapidity, so called $J/\psi$ puzzle at
RHIC, can well be explained by the competition between the two
production mechanisms. At LHC, however, the rapidity dependence of
the $J/\psi$ production is dominated by the regeneration process.
\end{abstract}

\date{\today}

\maketitle

From lattice quantum chromodynamics (QCD) calculations, a new
state of matter, the so-called quark-gluon plasma (QGP), has been
predicted to exist at high temperature and/or high  baryon
density. The $J/\psi$ production has long been considered as a
probe of the QGP formation in relativistic heavy ion
collisions~\cite{matsui}. Different from the $J/\psi$ suppression
observed at the CERN Super Proton Synchrotron (SPS) where almost
all the charmonia are primordially produced through hard
nucleon-nucleon collisions (NN) and then suffer from nuclear
absorption~\cite{gersche,vogt} in the initial stage and anomalous
suppression in the hot and dense
medium~\cite{sps,blaizot,polleri,capella,bratkovskaya,hufner} in
the later stage, there are a remarkable number of charm quarks in
the QGP produced in high energy nuclear collisions at the BNL
Relativistic Heavy Ion Collider (RHIC) and the CERN Large Hadron
Collider (LHC), and the regeneration, namely the continuous
recombination of those uncorrelated charm quarks offers another
source for $J/\psi$
production~\cite{pbm,gorenstein,grandchamp,greco,thews}.
Obviously, the regeneration will enhance the $J/\psi$ yield and
alter its momentum spectrum. At midrapidity, the initial
production and regeneration are almost equally important and the
competition between the two production mechanisms
controls~\cite{yan,liu} the $J/\psi$ nuclear modification factor
$R_\text{AA}$ and averaged transverse momentum square $\langle
p_t^2\rangle$ observed at RHIC~\cite{phenix1,star1}.

Recently, the rapidity dependence of $J/\psi$ production in Au+Au
collisions is measured at RHIC~\cite{phenix1,phenix2} and
discussed in models~\cite{kharzeev,andronic,zhao1}. As one can
see, as a function of the number of participant nucleons $N_p$,
the observed $R_{AA}(N_P)$ shows that the suppression is
significantly stronger at forward rapidity ($|y| \in [1.2, 2.2]$)
than that at midrapidity ($|y|<0.35$). The enhanced suppression is
especially clear in central collisions, see Figs. 4 and 5 of
Ref.~\cite{phenix1}. For $N_p >150$, the ratio of $R_{AA}^{\text
{forward}}/R_{AA}^{\text {mid}}$ is about $0.6$. In addition, we
observe that the value of $\langle p_t^2\rangle$ is lower at
midrapidity than that at forward rapidity, see Table II of
Ref.~\cite{phenix1}. The above observation is difficult to be
explained by models with only initial charmonium
production~\cite{phenix1}. Since the medium temperature at
midrapidity is higher than or at least the same as that at forward
rapidity, the suppression at forward rapidity is predicted in
these models to be less than or at most the same with that at
midrapidity.

In this Letter, we investigate the rapidity dependence of $J/\psi$
production by constructing a three dimensional transport equation
for the charmonium motion in the hot medium, with both initial
production and regeneration. When the regeneration mechanism is
included, the above $J/\psi$ puzzle is possible to be
explained~\cite{andronic}. Since the charm quarks are mainly
distributed in the central region, their recombination into
$J/\psi$s at midrapidity is more important than that at forward
rapidity. While the medium absorption at midrapidity is in
principle larger than that at forward rapidity, the competition
between the regeneration and absorption may lead to an enhancement
in $R_{AA}$ at midrapidity. We note that the effect in the
rapidity dependence of $R_{AA}$ should also be elaborated with the
rapidity dependence of $\langle p_t^2\rangle$. Considering the
fact that the measured forward rapidity is still located at the
plateau of light hadron rapidity distribution~\cite{brahms}, the
temperatures of the medium and in turn the $J/\psi$ absorption in
the mid and forward rapidity are approximately the same.
Therefore, the rapidity dependence of the regeneration becomes the
key factor to control the experimental observation
$R_\text{AA}^\text{forward}/R_\text{AA}^\text{mid} < 1$ at any
$N_p$ and $\langle p_t^2\rangle^\text{forward}/\langle
p_t^2\rangle^\text{mid}> 1$ at large $N_p$ at RHIC.

Since the experimental results can not separate primordial
$J/\psi$ and $J/\psi$ from $\chi_c$ and $\psi'$ decay, we consider
a transport equation for a charmonium distribution function
$f_\Psi(p,x)$ ($\Psi=J/\psi, \psi', \chi_c$). Using Lorentz
covariant time $\tau=\sqrt {t^2-z^2}$, space rapidity
$\eta=1/2\ln\left[(t+z)/(t-z)\right]$, transverse energy
$E_t^\Psi=\sqrt {E_\Psi^2-p_z^2}$ with $E_\Psi=\sqrt{m_\Psi^2+{\bf
p}^2}$ and momentum rapidity
$y_\Psi=1/2\ln\left[(E_\Psi+p_z)/(E_\Psi-p_z)\right]$, the three
dimensional transport equation can be expressed as
\begin{eqnarray}
\label{trans}
&&\left[\cosh(y_\Psi-\eta){\frac{\partial}{\partial\tau}}+{\frac{\sinh(y_\Psi-\eta)}{\tau}}{\frac{\partial}{\partial
\eta}}+{\bf
v}_t^\Psi\cdot\nabla_t\right]f_\Psi\nonumber\\
&=&-\alpha_\Psi f_\Psi+\beta_\Psi,
\end{eqnarray}
where ${\bf v}_t^\Psi={\bf p}_t/E_t^\Psi$ is the transverse
velocity which leads to the leakage effect and is proven to be
important for those high velocity charmonia at SPS~\cite{hufner},
and the suppression and regeneration in hot medium are described
by the loss term $\alpha_\Psi$ and gain term $\beta_\Psi$.
Considering the gluon dissociation process $\Psi + g \to c+\bar c$
and charmonium regeneration process $c+\bar c\to \Psi +g$ in QGP,
$\alpha\left({\bf p}_t,y_\Psi,{\bf x}_t,\eta,\tau\right)$ and
$\beta\left({\bf p}_t,y_\Psi,{\bf x}_t,\eta,\tau\right)$ are,
respectively, the momentum integration of the dissociation
probability $W_{g\Psi}^{c\bar c}$ multiplied by gluon thermal
distribution $f_g$ and the regeneration probability
$W_{c\bar{c}}^{g\Psi}$ multiplied by charm quark distribution
$f_c$~\cite{yan}. $W_{g\Psi}^{c\bar c}$ is calculated with
perturbative Coulomb potential~\cite{peskin} and
$W_{c\bar{c}}^{g\Psi}$ can be obtained from $W_{g\Psi}^{c\bar c}$
via detailed balance. Since the experimentally measured D-meson
flow is comparable with light hadron flow~\cite{phenix3}, we
assume charm quark thermalization and choose the density
$f_c(q,x)=T_A({\bf x}_t)T_B({\bf x}_t-{\bf b})\cosh\eta/\tau
d\sigma_{NN}^{c\bar c}/d\eta f_q(q,x)$, where $f_q$ is the
normalized charm quark thermal distribution, $d\sigma_{NN}^{c\bar
c}/d\eta$ the rapidity distribution of charm quark pairs produced
in NN collisions, ${\bf b}$ the impact parameter, and $T_A$ and
$T_B$ are the thickness functions of the two colliding nuclei
defined as $T({\bf x}_t)=\lim_{z_1\to -\infty, z_2\to\infty}T({\bf
x}_t,z_1,z_2)$ with $T({\bf
x}_t,z_1,z_2)=\int_{z_1}^{z_2}dz\rho({\bf x}_t,z)$ and the
Woods-Saxon nuclear density profile $\rho({\bf x})$.

Since the hadronic phase appears later in the evolution of the
fireball when the density of the system is lower compared to the
early hot and dense period, we have neglected the charmonium
production and suppression in hadron gas. On the other hand, from
the lattice QCD simulation~\cite{asakawa} on charmonium spectral
function at finite temperature, $J/\psi$ collapses at a
dissociation temperature $T_d^{J/\psi}$, which corresponds to the
idea of sequential suppression~\cite{satz}. Taking into account
these two aspects, the dissociation and regeneration processes
happen only in the temperature region $T_c<T<T_d^\Psi$ of  QGP,
where $T_c$ is the critical temperature of deconfinement phase
transition. Outside this region there are
$\alpha_\Psi=\beta_\Psi=0$ for $T<T_c$ and
$\alpha_\Psi=\beta_\Psi=\infty$ for $T>T_d^\Psi$.

The transport equation has been solved analytically and the result
is shown as
\begin{widetext}
\vspace{-7mm}
\begin{eqnarray}
\label{solution}
f_\Psi\left({\bf p}_t,y_\Psi,{\bf
x}_t,\eta,\tau\right)&=&f_\Psi\left({\bf p}_t,y_\Psi,{\bf
X}_\Psi(\tau_0),H_\Psi(\tau_0),\tau_0\right)e^{-\int^{\tau}_{\tau_0}d\tau'
\alpha_\Psi\left({\bf p}_t,y_\Psi,{\bf X}_\Psi(\tau'),H_\Psi(\tau'),\tau'\right)/\Delta(\tau')}\nonumber\\
&+&\int^{\tau}_{\tau_0}d\tau' \beta_\Psi\left({\bf
p}_t,y_\Psi,{\bf
X}_\Psi(\tau'),H_\Psi(\tau'),\tau'\right)/\Delta(\tau')\
e^{-\int^{\tau}_{\tau'}d\tau''\alpha_\Psi\left({\bf
p}_t,y_\Psi,{\bf
X}_\Psi(\tau''),H_\Psi(\tau''),\tau''\right)/\Delta(\tau'')}
\end{eqnarray}
\end{widetext}
with
\begin{eqnarray}
\label{xh}
&& {\bf X}_\Psi(\tau')={\bf x}_t-{\bf
v}_t^\Psi\left[\tau\cosh(y_\Psi-\eta)
-\tau'\Delta(\tau')\right],\nonumber\\
&& H_\Psi(\tau')=y_\Psi-\textrm{arcsinh}\left(\tau/\tau'\text{sinh}(y_\Psi-\eta)\right),\nonumber\\
&&
\Delta(\tau')=\sqrt{1+(\tau/\tau')^2\text{sinh}^2(y_\Psi-\eta)}.
\end{eqnarray}
The first and second terms on the right-hand side of the solution
(\ref{solution}) indicate the contributions from the initial
production and continuous regeneration, respectively, and both
suffer anomalous suppression in the medium. Since the collision
time for NN interactions at RHIC energy is about 0.1 fm/c and less
than the starting time $\tau_0$ of the medium evolution which is
about 0.5 fm/c, the nuclear absorption and Cronin
effect~\cite{gavin} for the initially produced charmonia have
ceased before the QGP evolution and can be reflected in the
initial distribution of the transport equation (\ref{trans}). From
the PHENIX d+Au data, the cold nuclear matter effect is smaller at
RHIC than those observed at lower energies~\cite{phenix4}.
Considering further a finite formation time of charmonia which is
about 0.5 fm/c and larger than the collision time, the nuclear
absorption can approximately be neglected for the calculations at
high energies~\cite{liu}. In this case, the initial distribution
$f_\Psi\left({\bf p}_t,y_\Psi,{\bf x}_t,\eta,\tau_0\right)$ is
calculated according to the Glauber model with the $J/\psi$
production cross section $d\sigma_{NN}^\Psi/(p_tdp_tdy_\Psi)$ in
NN collisions and the modification of the Cronin
effect~\cite{gavin}. The coordinate shifts ${\bf x}_t \to {\bf
X}_\Psi$ and $\eta \to H_\Psi$ in the solution (\ref{solution})
reflect the leakage effect in the transverse and longitudinal
directions.

The local temperature, baryon chemical potential and fluid
velocity are determined by hydrodynamic equations and control the
suppression and regeneration region through the gluon and quark
thermal distributions. Similar to the three dimensional transport
equation for $\Psi$ motion, a self-consistent treatment for the
calculation of $J/\psi$ rapidity dependence needs three
dimensional hydrodynamic equations. Considering the fact that the
experimentally observed forward rapidity $1.2< y <2.2$ is still
located at the central plateau of the Bjorken hydrodynamics at
RHIC energy~\cite{brahms}, we can, for simplification, neglect the
rapidity dependence of the fluid and take the transverse
hydrodynamic equations~\cite{yan,liu} at $y=0$ for the QGP
evolution at both mid and forward rapidity. To close the
equations, we take the equation of state~\cite{sollfrank} of ideal
gases of partons and hadrons with a first order phase transition
at $T_c$, and the initial condition for the hydrodynamics at RHIC
is the same as in Ref~\cite{yan,liu}.

We now fix the charmonium and charm quark distributions in NN
collisions. With the precise transverse momentum and rapidity
distributions for $J/\psi$ measured in proton-proton collisions at
RHIC energy~\cite{phenix5}, the distribution
$d\sigma_{NN}^\Psi/(p_tdp_tdy_\Psi)$ can be parameterized as
\begin{equation}
\label{ppjpsi}
{\frac{d\sigma_{NN}^\Psi}{p_tdp_tdy_\Psi}}={2(n-1)\over
D(y_\Psi)}(1+{\frac{p_t^2}{D(y_\Psi)}})^{-n}{\frac{d\sigma_{NN}^\Psi}{dy_\Psi}}
\end{equation}
with $n=6$, $D(y_\Psi)={\langle
p_t^2\rangle_{NN}(n-2)}(1-y_\Psi^2/Y_\Psi^2)$, the maximum
rapidity $Y_\Psi=\text{arccosh}(\sqrt{s_{NN}}/(2m_\Psi))$, the
averaged transverse momentum square $\langle
p_t^2\rangle_{NN}=4.14$ (GeV/c)$^2$ and a double Gaussian
distribution $d\sigma_{NN}^\Psi/dy_\Psi$~\cite{phenix5}. At LHC
energy, from the CDF experiment and CEM
calculation~\cite{alice,accardi}, we take $n=4$, $\langle
p_t^2\rangle_{NN}=12$ (GeV/c)$^2$ and
$d\sigma_{NN}^\Psi/dy_\Psi=2\ \mu b$. Taking into account the
Cronin effect~\cite{gavin}, $\langle p_t^2\rangle_{NN}$ is
replaced by $\langle p_t^2\rangle_N=\langle
p_t^2\rangle_{NN}+a(T_A({\bf x}_t,-\infty,z_A)+T_B({\bf x}_t-{\bf
b},z_B,\infty))$. The second term is from the multiply scattering
of the two gluons with nucleons before they fuse into a $\Psi$,
and the constant $a$ is adjusted to the data for pA
collisions~\cite{yan,zhao}.

For $d\sigma_{NN}^{c\bar c}/d\eta$ at RHIC, there is a large
experimental uncertainty and the difference among theoretical
model estimations is also significant~\cite{zhang}. We take a
Gaussian distribution $d\sigma_{NN}^{c\bar c}/d\eta =
d\sigma_{NN}^{c\bar c}/d\eta\big|_{\eta=0}e^{-\eta^2/\eta_0^2}$
with $d\sigma_{NN}^{c\bar c}/d\eta\big|_{\eta=0}=120\ \mu b$ which
agrees with the experimental data~\cite{zhang} and a parameter
$\eta_0$ determined by the assumption $(d\sigma_{NN}^{c\bar
c}/d\eta|_{\eta=1.7})\big/(d\sigma_{NN}^{c\bar
c}/d\eta|_{\eta=0})=1/3$ which is in between the maximum and
minimum model calculations~\cite{zhang}. At LHC, we take
$d\sigma_{NN}^{c\bar c}/d\eta$ directly from the PYTHIA
simulation~\cite{pythia}.

\begin{figure}[!htbp]
\includegraphics[width=8cm]{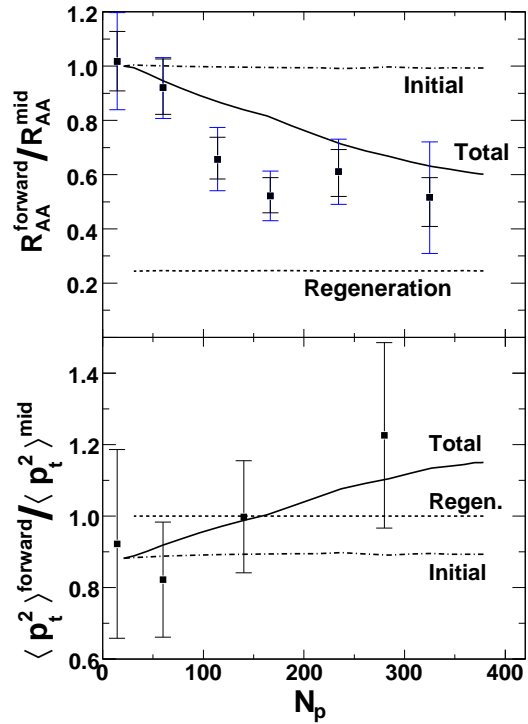}
\vspace{-5mm} \caption {The ratio of forward to mid rapidity
$R_{AA}$ and the ratio of forward to mid rapidity $\langle
p_t^2\rangle$ at RHIC. The lines are theoretical calculations and
the data are from \cite{phenix1,phenix2}. } \label{fig1}
\end{figure}

Taking the dissociation temperatures $T_d^{J/\psi}=1.9\ T_c,\
T_d^{\chi_c}=T_d^{\psi'}=T_c=165$ MeV and the $J/\psi$ fractions
from direct production and $\chi_c$ and $\psi'$ decay as $6:3:1$,
we integrated the $J/\psi$ distribution $f_{J/\psi}$ on the
hadronization hyper surface and calculated the $J/\psi$ $R_{AA}$
and $\langle p_t^2\rangle$ at mid and forward rapidity in heavy
ion collisions at RHIC and LHC energies. In the upper panel of
Fig.\ref{fig1} we show the ratio of forward to mid rapidity
$R_{AA}$ as a function of $N_p$ at RHIC. The new experimental
founding of $R_{AA}^\text{forward}/R_{AA}^\text{mid}<1$ is
difficult to be understood in models with only initial production
mechanism, since the suppression in forward region should be
smaller than or at most the same as that in central region. In our
calculation with the assumption of the same medium for the two
rapidity regions, the ratio with only initial production is almost
a constant, the small deviation is from the rapidity dependence of
the $J/\psi$ transverse momentum distribution in NN collisions,
see $D(y_\psi)$ in (\ref{ppjpsi}). For the case with only
regeneration, the ratio becomes a strict constant and is
determined by the cross sections $\sigma_\Psi$ and $\sigma_{c\bar
c}$ in NN collisions. Since the regeneration contribution at
forward rapidity is smaller than that at midrapidity, the total
ratio is less than the value with only initial production in the
whole $N_p$ region. However, it can not reach the regeneration
limit, because at RHIC the maximum fraction of regeneration is
only around $50\%$~\cite{yan}.
\begin{figure}[!htbp]
\includegraphics[width=8cm]{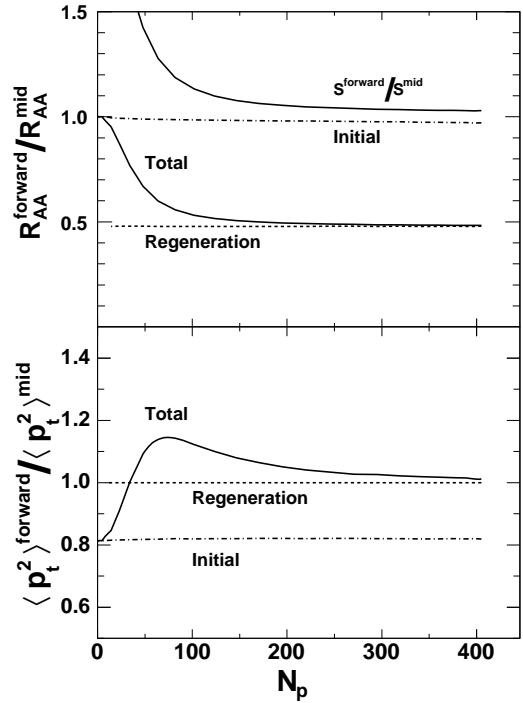}
\vspace{-5mm} \caption {The ratio of forward to mid rapidity
$R_{AA}$ , the ratio of forward to mid rapidity $\langle
p_t^2\rangle$ and the ratio of forward to mid rapidity $S_{AA}$ in
Pb+Pb collisions at LHC. } \label{fig2}
\end{figure}

The ratio of forward to mid rapidity $\langle p_t^2\rangle$ is
shown in the lower panel of Fig.\ref{fig1}. Again it is a constant
in the case with only initial production or regeneration. While
the $\langle p_t^2\rangle$ is dominated by the initial production
at forward rapidity, both initial production and regeneration
affect the midrapidity region. In addition, the initially produced
$J/\psi$s are harder in momentum distribution. Therefore, the
total ratio is always larger than the limiting value set by the
initial production. It starts with the value of $\langle
p_t^2\rangle_{NN}^\text{forward}/\langle
p_t^2\rangle_{NN}^\text{mid}$ and goes up with increasing $N_p$
monotonously.

While the $J/\psi$ production at RHIC is governed by the
competition between the initial production and regeneration, the
situation at LHC is very different. At LHC, the fireball is
hotter, larger, and longer lived, almost all the initially
produced $J/\psi$s are eaten up by the QGP. On the other hand,
there are about ten times more charm quarks at LHC than that at
RHIC. As a result, the regeneration becomes much more important
and dominates the $J/\psi$ production in heavy ion collisions at
LHC. We show our prediction for the two ratios in Fig.\ref{fig2}.
The midrapidity and forward rapidity regions are defined as
$|y|<0.9$ and $y\in [2.5,4]$. With increasing $N_p$, both ratios
start from the initial production limit and reaches the
regeneration limit fast. At large $N_p$, the two ratios are almost
fully controlled by the regeneration. The rapid change of the two
ratios at low $N_p$ is due to the important contribution from the
initial production in peripheral collisions, like the case in
central collisions at RHIC.

When the $J/\psi$ regeneration becomes dominant, we can use the
quantity $S_{AA}=N_{J/\psi}/N_D^2$ to describe the medium effect,
where $N_{J/\psi}$ and $N_D$ are, respectively, the $J/\psi$
number and D-meson number per unit rapidity. Since the number of
initially produced $J/\psi$s is proportional to the number of
binary collisions $N_c$ and the number of regenerated $J/\psi$s to
$N_c^2$, we have $S_{AA}\sim 1/N_c$ in the initial production
dominant region and $S_{AA}\sim 1$ in the regeneration dominant
region. The ratio of forward to mid rapidity $S_{AA}$ at LHC is
shown in Fig.\ref{fig2}. In contrast to the ratio of $R_{AA}$
which reaches unity in the limit of small $N_p$, the ratio of
$S_{AA}$ approaches to unity in the limit of large $N_p$.

In summary, we have constructed a three dimensional transport
model to describe the rapidity dependence of $J/\psi$ production
in relativistic heavy ion collisions. To understand the production
and suppression mechanisms, we need to investigate simultaneously
the $J/\psi$ yield and transverse momentum distribution. At RHIC,
the competition between the initial production and regeneration
explains well the rapidity dependence of $J/\psi$ $R_{AA}$ and
$\langle p_t^2\rangle$, namely
$R_{AA}^\text{forward}/R_{AA}^\text{mid} <1$ at any centrality and
$\langle p_t^2\rangle^\text{forward}/\langle
p_t^2\rangle^\text{mid}>1$ for semi-central and central
collisions. At LHC, the two ratios approach to the corresponding
regeneration limits already in semi-central collisions, due to the
fast dissociation of the initially produced charmonia and the
dominance of the regeneration.

{\bf Acknowledgement:} The work is supported by the NSFC Grant
10735040, the 973-project 2006CB921404 and 2007CB815000, and the
U.S. Department of Energy under Contract No. DE-AC03-76SF00098.

\end{document}